\documentstyle[prl,multicol,aps]{revtex}
\begin{document}
\draft
\title{Properties of Particles Obeying Ambiguous Statistics}
\author{M.V.~Medvedev \cite{email,mm} 
 }
%\thanks{ Also:
%{Russian Research Center ``Kurchatov Institute", Institute for Nuclear 
%Fusion, Moscow 123182, RUSSIA.}}
%{ } and  P.H.~Diamond \cite{pd}
%\thanks{ Also:
%{General Atomics, San Diego, California 92122.}}
%}
\address{Physics Department, University of California at San Diego,
La Jolla, California 92093.}
%\\[5pt]
%\em Physics Department, University of California at San Diego\\
%\em La Jolla, California 92093.
%\date{\today}
\maketitle

\begin{abstract}
A new class of identical particles which may exhibit both Bose and Fermi 
statistics with respective probabilities $p_b$ and $p_f$ is introduced. Such an
uncertainity may be either an intrinsic property of a particle or can be 
viewed as an ``experimental uncertainity''.
Statistical equivalence of such particles and particles obeying 
parastatistics of infinite order is shown. Generalized statistical 
distributions are derived and
statistical and thermodynamical properties
of an ideal gas of the particles are investigated. The physical nature of such
particles and the implications of this investigation for the statistics of 
extremal black holes are discussed.
\end{abstract}
\pacs{05.30.-d, 04.70.Dy, 14.80.Pb, 05.70.Ce}
%\newpage
%\baselineskip18pt

%\twocolumn
%\narrowtext
%\section{Introduction}
%\label{sec:intro}

\begin{multicols}{2}
Historically, the first attempt to generalize quantum Bose and Fermi statistics 
was made by Gentile \cite{Gentile}, who proposed  statistics in which up to $k$
particles are allowed to occupy a single quantum state (instead of one for
Fermi and infinitely many for Bose cases). Further generalizations were 
mostly on deformations of commutation relations for
particle creation and annihilation operators 
\cite{Green!Greenberg-book!Greenberg64,Govorkov,Greenberg90,Greenberg91}.
A special case (the so called ``infinite'' statistics) is rather interesting
in that the particles which obey this statistics
have nonnegative square norms \cite{Mohapatra90} for $|q|^2\le 1$,
while the observables have nonlocal properties. The commutation relations 
are simply \cite{Greenberg90,Greenberg91}:
\begin{equation}
a_i^{ }a_j^\dag-qa_j^\dag a_i^{ }=\delta_{ij} ,
\label{q-comm}
\end{equation}
where $q$ is either a real or complex parameter, $a_j^\dag$  and  $a_j$
are the particle creation and annihilation operators, and $\delta_{ij}$ is the
Kronecker's delta. 
Stimulated by recent
experimental observation of the {\em fractional quantum Hall effect}
\cite{Stromer82!Stromer83} in a two-dimensional electron gas, various
nontraditional statistics have been proposed
\cite{Leinaas&Myrheim!Wilczek,Halperin!Laughlin83!Calogero!Sutherland,%
Murthy&Shankar,Haldane91}. Most of them are based on
generalizations of particle wave-function permutation symmetries.
The concept of anyons, which are charge carriers in the fractional quantum 
Hall effect, is essentially two-dimensional \cite{Leinaas&Myrheim!Wilczek}.
However, in the strong magnetic field with fluxes antiparallel to the
ambient field and in the lowest Landau level, anyons are shown
\cite{Das!Wu} to obey the one-dimensional (though valid in any dimensions) 
Haldane's exclusion principle \cite{Haldane91}.
At last, the equivalence of the anyon statistics (in the above case) 
and $q$-on statistics, Eq.\ (\ref{q-comm}), with $q=e^{i\alpha\pi}$
($\alpha$ is the meassure of Haldane's/Pauli blocking and $\alpha\pi$
is exactly the wave-function phase-shift due to permutation of any two
particles) was proven \cite{Goldin&Sharp} via the properties of the
$N$-anyon permutation group.

Statistical and thermodynamical properties of systems of particles obeying 
fractional statistics (i.e., complex $q$ such as $|q|^2=1$) were 
extensively studied in the last several
years \cite{Murthy&Shankar,Das!Wu,Isakov!Sen&Bhaduri!Smezi}, whereas
these properties of particles with real $q$, $-1 < q < 1$, 
were not investigated (with some special exceptions
\cite{Inomata,Werner}). One should mention a special case
of the infinite statistics
\cite{Greenberg90} with $q=0$ which was suggested \cite{Govorkov} 
to be equivalent to the statistics of nonidentical particles
(quantum Boltzmann statistics). One should note here that it was also
recently suggested \cite{Strominger} that extremal black holes should 
obey such a statistics. This issue will be discussed below.

In this work we investigate a system of particles obeying ambiguous 
statistics which are (statistically) equivalent to those obeying 
the $q$-deformed statistics of real $q$. The partition function of 
a free gas of such particles, however, differs from that of a free 
$q$-on gas. The last is shown \cite{Werner} to be independent of the deformation
parameter $q$. In fact, the second quantized approach used in 
\cite{Werner} does not take into account the dynamics
of the internal degrees of freedom. Thus, the $q$-ons are always
distinguishable particles. Our model, however, is (probably) identical to 
the case when the internal degrees of freedom are in the thermodynamic 
equilibrium with the external degrees of freedom
(at least for $q=0$), and thus the particles are {\em not completely
distinguishable}. (Note, the particles are indistinguishable provided they
are in the same internal state, even if they have infinite number of internal
degrees of freedom.) Consequently, the quantum Boltzmann
distribution ($q=0$) is different from the classical Boltzmann.

Unlike the theories mentioned above, we admit only ``primary'' Bose-Einstein
and Fermi-Dirac statistics as  existing. Assume now that a particle
is neither a pure boson or pure fermion. Let another particle, which interacts 
with the first one, plays a role of an external observer. 
During the interaction, it performs a measurement at the first particle 
and identifies it as either a boson or a fermion with respective probabilities
$p_b$ and $p_f$. According to the result of this measurement, it interacts with
the first particle as if the last is a boson or fermion, respectively.
The first particle, of course, is the observer for the second
particle, thus the process is symmetric. 
Note, that $p_b+p_f$ is not necessarily equal to one,
and if not, it means that the second particle (observer)
does not detect the first particle. The probability of this is $1-p_b-p_f$.
The ``statistical unsertainity'' introduced above may be either the intrinsic 
property of a particle itself or the ``experimental unsertainity'' of the 
meassurement process. This  concept of ``imperfect measurement'' is similar 
to that proposed in order to resolve the causality paradox
in superluminal particle (tachyon) systems \cite{Schulman}. 
This model can be interpreted in another manner. Assume a particle can
oscillate between two types of statistics, then the model we propose represents
a system of such particles averaged over time-scales much larger than the 
oscillation period. The probabilities  $p_b$ and $p_f$, thus, are those portions
of time during which a particle resides in a Fermi- or Bose-type state. The 
model of this kind is relevant to systems of elementary particles which have
mutual supersymmetrical  partners (e.g., a photon and photino, etc.) when
transitions between these supersymmetrical states can occur 
\cite{Kane}. With some modifications, this model can also
be applied to systems of particles with changing flavor, e.g., quarks, gluons,
neutrinos, etc.

There is a conjectured relation between the particles of the ambiguous 
statistical type
and the $q$-deformed infinite statistics. Let $b_j^{ }$ and $b_j^{\dag}$ 
are the annihilation and creation operators for such a particle. The particle
exhibit bosonic properties with the probability $p_b$ and fernionic ones with 
$p_f$. Thus, a bilinear commutation relation for $b_j^{ }$ and $b_j^{\dag}$ 
is of boson type with the probability $p_b^2$, of fermion type with the 
probability $p_f^2$, and the two particles are nonidentical with the probability 
$2p_b p_f$, that is interchange of their positions result in
 another wave-function, i.e., $b_i^{ }b_j^{\dag}=\delta_{ij}$. We write 
\begin{eqnarray}
(p_b^2+p_f^2+2p_b p_f)b_i^{ }b_j^{\dag}
&-&(p_b^2-p_f^2)b_j^{\dag}b_i^{ } \nonumber\\ 
& &{ }=(p_b^2+p_f^2+2p_b p_f)\delta_{ij} .
\end{eqnarray}
This equation coinsides with Eq.\ (\ref{q-comm}), and the deformation
parameter is $q=(p_b-p_f)/(p_b+p_f)$. In the rest of this letter 
we derive statistical distributions for identical particles which
have ambiguous quantum exclusion properties and investigate some
thermodynamical properties of an ideal gas of such particles.

 A grand partition function for a system of particles with properties
defined by a stochastic label with a known probability distribution is a sum 
over all possible realizations (with a weight factor) of the partition functions
corresponding to each realization. In each realization, the system effectively 
consists of $k$ bosons and $N-k$ fermions ($N$ is the total number of
particles), while the probability of this realization is $p_b^k p_f^{N-k}$.
The total number of states of the system is
\begin{equation}
W=\prod_j\ \sum_{k=0}^{N_j} {N_j \choose k}
%\biggl(\begin{array}{c}\!\! N_j
%%\\[-0.04pc] 
%\\[-7.5pt]
%\!\! k \end{array}\!\! \biggr)
{w_b}_j(k) {w_f}_j(N_j-k) p_b^k p_f^{N_j-k} ,
\label{W}
\end{equation}
 where the weight factor 
%$\biggl(\begin{array}{c}\!\! N
%%\\[-0.04pc] 
%\\[-7.5pt]
%\!\! k \end{array}\!\! \biggr)
%\displaystyle{=\frac{N!}{k!(N-k)!}}$ 
${N\choose k}=\frac{N!}{k!(N-k)!}$ represents the number of
ways to arrange $N$ identical particles in two groups of respectively
$k$ and $N-k$ particles each. Here $w_b(m)$ and $w_f(m)$ are the numbers
of quantum states of $m$ identical particles occupying a group of $G$ states,
respectively, for bosons and fermions,
\begin{equation}
w_b(m)=\frac{(G+m-1)!}{m!(G-1)!}\ \ \hbox{and}\ \ 
w_f(m)=\frac{G!}{m!(G-m)!} .
\label{b-f}
\end{equation}
The partial partition function (expression under the product sign) can
be simplified and rewritten in terms of a generalized hypergeometric function
\begin{mathletters}
\label{Wj}
\begin{eqnarray}
W_j&=&p_f^{N_j}\frac{G_j}{N_j}\sum_{k=0}^{N_j}
{N_j\choose k}^2 {G_j-1+k\choose N_j-1}
%\biggl(\begin{array}{c}\!\! N_j
%%\\[-0.04pc] 
%\\[-7.5pt]
%\!\! k \end{array}\!\! \biggr)^2
%\biggl(\begin{array}{c}\! G_j-1+k
%%\\[-0.04pc] 
%\\[-7.5pt]
%\! N_j-1 \end{array}\! \biggr)
\left(\frac{p_b}{p_f}\right)^k 
\label{Wja}\\
&=&p_f^{N_j}G_j\frac{G_j!}{N_j!(G_j-N_j+1)!}\ \nonumber\\
& &{ }\cdot {_3F_2}\left(-N_j,-N_j,G_j;1,G_j-N_j+1;\frac{p_b}{p_f}\right) .
\label{Wjb}
\end{eqnarray}
\end{mathletters}

To proceed further, we assume the number of particles in a system to be
very large, and use Stirling approximation for the factorials and
Laplace's approximation method to evaluate the sum.
It is enough to keep the leading term, only.
Then one can write Eq.\ (\ref{Wj}) as follows
\begin{equation}
W_j\simeq {N_j\choose {k_0}_j}^2 {G_j+{k_0}_j\choose N_j}
%\biggl(\begin{array}{c}\! N_j
%%\\[-0.04pc] 
%\\[-7.5pt]
%\! {k_0}_j \end{array}\! \biggr)^2
%\biggl(\begin{array}{c}\! G_j+{k_0}_j
%%\\[-0.04pc] 
%\\[-7.5pt]
%\! N_j \end{array}\! \biggr)
p_b^{{k_0}_j} p_f^{N_j-{k_0}_j} ,
\label{Wj-appr}
\end{equation}
where ${k_0}_j$ is the solution of the equation
\begin{equation}
\left(\frac{N_j-{k_0}_j}{{k_0}_j}\right)^2
\left(\frac{G_j+{k_0}_j}{G_j-N_j+{k_0}_j}\right)\frac{p_b}{p_f}=1 .
\label{condition}
\end{equation}
Thus one can infer that the partition function, although defined by all
realizations, has its major contribution from that realization in which exactly 
$k_0$  bosons and $N-k_0$ fermions effectively exist, where $k_0$ is defined 
by the probability ratio $p_b/p_f$ alone. (Note, it is not a mixture of species, 
there is only one physically existing spesies.)  The entropy of the system is, 
as usual,
$S=\ln{W}$:
\begin{eqnarray}
S&\simeq&\sum_jG_j
\Bigl\{x_j\ln(p_b/p_f)+n_j\ln{p_f}+(x_j+1)\ln(x_j+1)
\nonumber\\
& &{ }+n_j\ln n_j-(1-n_j+x_j)\ln(1-n_j+x_j)
\nonumber\\
& &{ }-2\left[x_j\ln x_j+(n_j-x_j)\ln(n_j-x_j)\right]
\Bigr\} ,
\label{S}
\end{eqnarray}
where $n_j=N_j/G_j$ is the occupation number and $x_j={k_0}_j/G_j$.
Other thermodynamic functions follow straightforwardly. The most probable
distribution subject to the constraints that the total number of particles 
and the total 
energy  of the system be conserved, is determined by the condition
\begin{equation}
\frac{\partial}{\partial N_j}\left[S-\alpha\sum_j N_j
-\beta\sum_j\epsilon_j N_j\right]=0 .
\end{equation}
From this equation and Eqs.\ (\ref{condition}), (\ref{S}), we arrive at
\begin{equation}
\frac{n_j(1-n_j+x_j)}{(n_j-x_j)^2}p_f=\rho ,
\label{extremum}
\end{equation}
where $\rho=\exp\{(\epsilon_j-\mu)/T\}$, $\mu\equiv-\alpha T$ is the
chemical potential, and $T\equiv1/\beta$ is the temperature of the gas.
Equations (\ref{condition}) and (\ref{extremum}) define the occupation 
number $n(\epsilon_j)$ as a function of energy. Upon simple mathematical
manipulations, we obtain
\begin{equation}
x_j=\frac{n_j(\rho+\delta)-\delta}{2\rho+\delta}
\label{x}
\end{equation}
and
\begin{equation}
(\rho+\delta)(\rho+p_f)(\rho-p_b)n_j^2-2\rho(\rho+\delta)\sigma n_j
+\rho\delta^2=0 ,
\label{n-eq}
\end{equation}
where $\delta=p_f-p_b$ and $\sigma=p_f+p_b$. Note, the fraction
$x_j/n_j$ is the number of bosons in the system with respect to the 
total number of particles, i.e., $0\le x_j/n_j\le 1$.
The last equation can be explicitly solved to obtain
\begin{equation}
n_j={\sigma\rho\over(\rho+p_f)(\rho-p_b)}\left[1+\sqrt{1-
\frac{\delta^2}{\sigma^2}\frac{(\rho+p_f)(\rho-p_b)}{\rho(\rho+\delta)}}
\ \right].
\label{n}
\end{equation}
Here we omit the negative sign of the root, because it does not satisfy 
[together with Eq.\ (\ref{x})] the condition $0\le x_j/n_j\le 1$. 
In order to $n_j$ be positive number for all values of $\epsilon_j$,
the chemical potential should not exceed its maximum value
$\mu_{max}=-T\ln{p_b}$.
Eq.\ (\ref{n})
is a generalized energy distribution for the class of particles obeying 
the ambiguous
statistics as a function of two parameters $p_b$ and $p_f$ which define
the deviation from ``primary'' Bose or Fermi statistics. The generalized
distribution recovers Bose-Einstein and Fermi-Dirac ones in the limiting
cases 
%\hrule
$p_b=1, p_f=0$ and $p_b=0, p_f=1$, respectively. The case 
$p_b=p_f=\sigma/2$ is that of the quantum Boltzmann statistics since the
deformation parameter $q=0$. In this case
\begin{equation}
n_j=2\sigma\left(\rho-\frac{\sigma^2}{4}\frac{1}{\rho}\right)^{-1} ,
\label{gB}
\end{equation}
which is a quantum analog of the Boltzmann distribution. It is remarkable that
although particles can, in principle, be distinguished by their internal
states (they have an infinite number of internal degrees of freedom),
the  occupation number of these states is also energy and temperature 
dependent. Thus, the particles, in fact, are {\em not completely} 
distinguishable. Hence, some trend towards the familiar statistical (and 
thermodynamical) properties of identical particles may be discerned.
It is also interesting that Eq.\ (\ref{gB}) reduces to the classical Boltzmann
distribution $\sim\rho^{-1}$ for a weakly interacting gas, 
$\sigma\to 0$ (remember, particles
do not ``feel'' each other with the probability $1-p_b-p_f=1-\sigma$). The 
factor $2\sigma$ is not significant since it just renormalizes $\mu$.

All thermodynamic properties can be derived straightforwardly 
\cite{Landafshitz}. The chemical potential $\mu$ is defined implicitly by 
the equation for the total particle number
\begin{equation}
N=\frac{Vm^{2/3}}{\sqrt{2}\pi^2\hbar^3}\int_0^\infty\sqrt{\epsilon}\,
n(\epsilon)\,d\epsilon ,
\label{N}
\end{equation}
where $V$ is the volume of the gas, $m$ is the particle mass, $\hbar$ is the
Plank's constant. Then the equation of state can be found in a parametric form
from the thermodynamic potential $\Omega=-PV$,
\begin{equation}
\Omega=-\frac{2}{3}\frac{Vm^{2/3}}{\sqrt{2}\pi^2\hbar^3}\int_0^\infty
\epsilon^{3/2}\,n(\epsilon)\,d\epsilon ,
\label{Omega}
\end{equation}
In the classical Boltzmann limit $\exp\{\mu/T\}\ll 1$ the equation of state
\cite{Landafshitz} expanded to second order is  
\end{multicols}
\begin{eqnarray}
PV&=&NT\left\{1+
\frac{\delta}{\left(\sigma+2\sqrt{p_b p_f}\right)}\frac{\pi^{3/2}}{2}
\frac{N\hbar^3}{V(mT)^{3/2}}
-\left[\frac{(\delta^2 +p_b p_f)}{\left(\sigma+2\sqrt{p_bp_f}\right)^2}
+\frac{\delta^2\sqrt{p_b p_f}}{4\left(\sigma+2\sqrt{p_bp_f}\right)^3}\,\right]
\frac{16\pi^3}{3^{5/2}}\frac{N^2\hbar^6}{V^2(mT)^3}\right\} .
\label{eq-state}
\end{eqnarray}
\begin{multicols}{2}
There is an effective attraction between particles of the gas when $p_b>p_f$
and effective repulsion when $p_b<p_f$ (the second term overcomes the last 
one), as expected. In the case of the quantum Boltzmann statistics, $\delta=0$, 
the second term vanishes, but a residual attraction still exists (the last 
quadratic term).

As was mentioned above, extremal black holes are shown to obey the
infinite statistics with the deformation parameter $q=0$
\cite{Strominger}. Let's consider
a (``gedanken'') gas of charged, nonrotating, extremal black holes, and
assume their charges are of the same sign. Then, by definition,
$M^2=Q^2$ (with $M$ and $Q$ being the mass and charge of a black hole).
In the limit of large black hole separation (i.e., low-density, ideal gas
approximation, no relativistic effects), gravitational attraction of
``particles''  completely cancels electrostatic repulsion. Thus, the
system at hand is a gas of noninteracting particles (neutrals).
The thermodynamical properties of such a gas are governed by the
particle statistics properties, alone. According to Eq.\
(\ref{eq-state}) (with $p_b=p_f=\sigma/2, \delta=0$), the nonrotating, 
extremal black holes will experience
{\em weak statistical attraction}. Hence, one may expect that black
holes will form clusters, for instance, the ultimate size of which is 
controlled by the gas temperature (i.e., the velocity dispersion of black
holes) and completely independent of {\em any} black hole parameters
(e.g., their masses and charges).

To summarize, we introduced a new class of particles which may fluctuate
between Bose- and Fermi-type statistics or may have uncertainity in their 
statistics. A generalized distribution function 
is derived. For a particular case of the quantum Boltzmann statistics, the
distribution function (i.e., occupation numbers) is different from the classical
Boltzmann case. An equation of state in the classical Boltzmann limit is
derived. The relation of this type of statistics to  the existing 
generalizations of Bose and Fermi statistics has been ascertained.  
The implication of these results obtained to the ensemble of extremal black
holes is discussed.

I am grateful to P.H.~Diamond for useful discussions. 
This work was supported by DoE grant No. DEFG0388ER53275.

%\bibliography{ }
%
%\begin{figure}
%\caption{ }
%\label{fig1}
%\end{figure}
%
\end{multicols}
\end{document}